\documentclass[aps,12pt,preprint,groupedaddress,showpacs,reprintnumbers,amsmath,amssymb,floatfix]{revtex4}

\usepackage{graphicx}
\usepackage{dcolumn}
\usepackage{bm}
\usepackage{color}
\usepackage{rotating}

\begin{document}

\title{The outer crust of a cold, non-accreting neutron 
star \\within the Quark-Meson Coupling (QMC) model}


\author{S. Anti\'{c}}
 \email{sofija.antic@adelaide.edu.au}
 \affiliation{%
 CSSM and CoEPP, School of Physical Sciences, University of Adelaide, 
Adelaide SA 5005, Australia
 } 

\author{J. R. Stone}
\affiliation{%
Department of Physics and Astronomy, University of Tennessee, TN 37996 USA\\
Department of Physics (Astrophysics), University of Oxford, 
Oxford OX1 3RH, United Kingdom
}%
\author{J. C. Miller}
\affiliation{%
Department of Physics (Astrophysics), University of Oxford, 
Oxford OX1 3RH, United Kingdom
}%

\author{K. L. Martinez}
\affiliation{%
 CSSM and CoEPP, School of Physical Sciences, University of Adelaide, 
Adelaide SA 5005, Australia
}%

\author{A. W. Thomas}
\affiliation{%
CSSM and CoEPP, School of Physical Sciences, University of Adelaide, 
Adelaide SA 5005, Australia
}%
\author{P. A. M. Guichon}
\affiliation{%
IRFU-CEA, Universit\'e Paris-Saclay, F91191 Gif sur Yvette, France
}

\date{\today}

\begin{abstract}

The outer crust properties of cold non-accreting neutron stars are studied within the framework of the quark-meson coupling (QMC) model, which includes the effects of modifications of the quark structure inside individual nucleons when they are within a high-density nuclear medium. With a unique set of five well-constrained adjustable parameters, which have a clear physical basis, the QMC model gives predictions for the ground state observables of even-even nuclei which agree with experiment as well as traditional models. Furthermore, it gives improved theoretical values for nuclei thought to play a role in the outer crusts of neutron stars but for which experimental data is not available. Using the latest experimental data tables wherever possible but otherwise the predictions from the QMC model, we construct an equation of state for the outer crust which is then used within stellar model calculations to obtain an equilibrium sequence of crustal layers, each characterized by a particular neutron rich nuclei. Various properties of the layers are calculated for a range of neutron-star masses, and comparisons are made with alternative equations 
of state from the literature. This leads to the conclusion that the QMC model successfully predicts the outer crust properties and is fully comparable with the more traditional mass models, which all depend on a larger number of parameters.

\end{abstract}

\pacs{Valid PACS appear here}
\maketitle


\section{Introduction}
\label{sec:introduction}

Neutron stars (NSs) are the only known objects in the Universe that 
contain nuclear matter in equilibrium at densities as high as several 
times the saturation density $n_0$ ( $\sim 0.16$ fm$^{-3}$). These complex 
objects connect many different fields of research, from nuclear and 
particle physics and astrophysics to general relativity. Small in size,  
with radii less than $\sim 14$ km but having substantial mass (as high 
as two solar masses), they possess many extreme properties such as 
surface magnetic fields of up to $10^{13}$ Gauss (for a ``classical'' 
pulsar), rotation periods between $1.4$ ms - $30$ s, densities up to 
$10$ $n_0$ and gravity which can be $\sim 10^{11}$ times stronger than 
that of the Earth. There is still much about them which is not very well 
known, starting with the building blocks of heavy NS cores, the EOS of 
high density matter and the relation between NS masses and radii, going 
on to their thermodynamics and shape oscillations, magnetic fields and 
much more. However, progress in theoretical and terrestrial experimental 
efforts and the recent detections of gravitational waves from neutron 
star mergers, have put the research of NSs in stronger focus that ever 
before.

A neutron star is born in a core-collapse supernova event, at the end of the life of a massive star \cite{Woosley2005}. The core of the progenitor star, containing mainly iron-group nuclei, is rapidly transformed by subsequent electron captures and photodisintegration to a hot, very neutron rich object having about 10 -15 km radius and with most of its mass (1 - 2 M$_\odot$) concentrated in the center, the proto-neutron star. This star is initially fully fluid and has temperature $\sim 10^{10-11}$K \cite{yakovlev2001,Chamel2015}. The matter cools by neutrino emission and after a few years it reaches 
temperature $\sim 10^9$ K, allowing the outer layer of the star to form a solid crust beneath the envelope, an ocean of a hot Coulomb fluid plasma and a thin atmosphere (see e.g. 
\cite{Chamel2008,Chamel2015,Meisel2018,Fantina2020} and references therein). The matter stratifies into layers, each of them containing fully ionized atoms, arranged in a body centered cubic (bcc) lattice to minimize their Coulomb interaction energy, and relativistic, fully degenerate electrons, cooling further to temperatures of $\sim 10^6 - 10^7$ K. 

Moving inward from the NS surface, with growing pressure the electron chemical potential increases and electron capture by nuclei, $A(Z,N)+e\rightarrow A(Z-1,N+1)+\nu$, followed by neutrino emission becomes energetically favorable. The nuclei building different layers of the outer crust become more neutron rich with decreasing neutron separation energy until neutrons start to drip out. This process leads to the appearance of a free neutron gas as part of the background alongside the electrons, marking the transition to the NS inner crust. Further increase of density eventually causes nuclei to deform, touch and clump, transforming the lattice structure into various exotic shapes, called nuclear pasta. The nuclei are stabilized against $\beta$ decay by the filled electron Fermi sea. Eventually the matter goes through a phase 
transition to a homogeneous liquid mixture of neutrons, protons and electrons (and muons at higher densities), forming the core of an average mass NS. For a very heavy NS with mass close to $2 M_\odot$, the core structure may be divided into two separate regions: the homogeneous outer core consisting of n, p, e$^{-}$ and $\mu^{-}$, and the inner core with a density several times the saturation density of nuclear matter. Exotic particles such as hyperons and/or various phases of quark matter have been proposed to appear under these extreme conditions \cite{RevModPhys.89.015007}.

Theoretical models of the outer crust depend on only one nuclear physics input, the nuclear masses. The models evolve as improvements are made in experimental mass determinations and in the theoretical predictions for those nuclei which are too close to the neutron drip-line for measurements to be made. Feynman, Metropolis, and Teller \cite{Feynman1949} calculated the EOS of the envelope and 
established its ground state as being $^{56}Fe$. The outer crust EOS was first studied by Baym {\textit et.al} \cite{Baym:1971pw} in 1971, using the droplet model by Myers and Swiatecki \cite{Myers1966}. Haensel, Zdunik, and Dobaczewski \cite{Haensel1989} employed the Hartree-Fock-Bogolyubov formalism with the SkP Skyrme force and the Myers droplet model \cite{Myers1977}. Haensel and Pichon \cite{Haensel1994} used the experimental nuclear data mass table of 1992 from Audi and Wapstra \cite{Audi1993} and the theoretical nuclear mass tables of the droplet models from Moller and Nix \cite{Moeller1988} and Aboussir \textit{et al.} \cite{Aboussir1992}. R{\"u}ster \textit{et al.} \cite{PhysRevC.73.035804} updated the work of Baym using nuclear data available in 2006 and theoretical mass tables via the Brussels Nuclear Library for Astrophysics Applications (BRUSLIB) \cite{BRUSLIB} and by Dobaczewski and co-workers \cite{Dobaczewski2006} for Skyrme-based models, and by Geng, Toki, and Meng for a relativistic model \cite{Geng2005} (for details see \cite{PhysRevC.73.035804}). Pearson \textit{et al.} \cite{PhysRevC.83.065810} used the Hartree-Fock-Bogoliubov method with three Skyrme-force models (HFB-19, HFB-20, and HFB-21) and the Gogny-force model D1M. 

Chamel and Fantina \cite{Chamel2016} examined the validity of the generally accepted assumption that the layers in the outer crust each consist of a pure body-centered cubic ionic crystal in a charge compensating background of highly degenerate electrons. They studied the 
stability of binary and ternary compounds in different cubic and 
non-cubic lattices in dense stellar matter and showed that their 
stability against phase separation is uniquely determined by their structure and composition,  irrespective of the stellar conditions. In addition, they obtained the EOS and the ground-state structure for the outer crust of a non-accreting cold neutron star using the experimental 2012 Atomic Mass Evaluation \cite{Wang2012} and the HBF-24 \cite{Goriely2013} theoretical model for nuclear masses. Pearson \textit{et al.} \cite{Pearson2018,Pearson2019} constructed a unified EOS for NSs with theHFB-22, HFB-24 and HFB-26 models and studied the impact of the nuclear symmetry energy on properties of the NS interior. The question of crystallization in the outer crust of non-accreting and accreting cold neutron stars was further pursued by Fantina \textit{et al.} \cite{Fantina2020}. They concluded that the presence of impurities in the outer crust is non-negligible and may have a sizable impact on transport properties \cite{Schmitt2018}, which could be important for 
the cooling of neutron stars and their magneto-rotational evolution \cite{Gourgouliatos2018}.

In this work we study the outermost solid layer of a NS, the outer crust, for an isolated, non-magnetized, non-accreting NS in the framework of the QMC model. The results are compared with the outcome of calculations using the the finite range droplet (FRDM) model \cite{MOLLER20161}, the non-relativistic energy density functional (EDF) model with the Skyrme interaction - the HFB-24 
\cite{Goriely2013,Chamel2016,F.Fantina:2017xnk,Pearson2018} and the Walecka type relativistic mean-field (RMF) model with NL3 interaction \cite{LALAZISSIS19991}. The general conclusion of previous models of the outer crust under the same conditions has been that the EOS is rather insensitive to the nuclear model of theoretical masses, but the details of the layers and their composition depend on the predictive power of masses close to the neutron drip line. The experimental verification of theoretically predicted masses from different models is waiting for a new generation of terrestrial facilities and observational techniques. However, it is interesting to catalog the current results and add the new data from the QMC model to the existing predictions.

This paper is organised as follows: the QMC model is briefly introduced in Sec.~\ref{sec:QMC} with references to its development and applications, including the latest version, QMC$\pi$-III, used in this work to calculate unknown masses of neutron rich nuclei close to the neutron drip line. This is followed by description of the calculation method (Sec.~\ref{sec:OCcalc}) and a brief comment on theoretical models used for a comparison with the present QMC results (Sec.~\ref{sec:models}). The main results are presented in Sec.~\ref{sec:Results}, including the low density QMC$\pi$-III EOS, the sequence of Z and N for nuclei 
building the NS outer crust, the position of the neutron drip line and the size and content of individual layers of the outer crust. The gravitational and baryonic masses, and the electron density, Fermi momentum and specific heat are also calculated for each layer. In Sec.~\ref{sec:Conclusion} we summarize the outcome of the present work and discuss the future development of applying the QMC model to the outermost layers of NSs.

\section{Method}
\label{sec:methods}

\subsection{Theoretical framework of the QMC model}\label{sec:QMC}

One of the main goals of applying the QMC model to properties of neutron 
stars was to construct the QMC EOS over its full range of densities, 
covering both the core and the crust. The model has already been 
successfully applied to NS cores and has given NS masses as high as 
$\sim 2 M_{\odot}$ \cite{RIKOVSKASTONE2007341,Whittenbury2014}, even in 
the presence of hyperons at high densities. A somewhat different version 
of the QMC model has also been applied to high density matter in NSs 
exploring compatibility with GW170817 data \cite{Motta:2019tjc}. 
However, application of the QMC model to the NS crust has not been carried out until 
now. We address the outer crust in this work, leaving 
the inner crust, including the pasta region, to future publications.

The QMC model, proposed in 1988 \cite{GUICHON1988235} and developed extensively
over the following decades~\cite{Guichon:1995ue,Saito:1994ki,Saito:2005rv}, 
takes into account the internal quark structure of a nucleon in contrast to most of 
the traditional mean-field nuclear structure models, which consider the 
nucleons as point-like objects. When the nucleon is immersed in a mean 
field created by surrounding nucleons, the effects of the external 
fields is self-consistently related to the dynamics of the quarks in the 
nucleon. In the present version of the QMC model the light quark 
confinement in a nucleon is schematically modeled with an MIT bag and 
the interaction between the quarks in individual bags is described by 
the exchange of effective mesons ($\sigma$, $\omega$, $\rho$ and $\pi$). 
It is found that the application of the scalar mean field $\sigma$, 
with strength up to a half of the nucleon mass, can lead to 
significant changes in the structure of the nucleon. Solution of the bag model 
equations of motion in a constant scalar field yields an 
effective mass
\begin{equation} 
M^{\ast}_B = M_B - g_{\sigma}\sigma + \frac{d}{2}(g_{\sigma}\sigma)^2 \, , 
\end{equation} 
with $g_{\sigma}\sigma$ being the strength of the scalar field. Here 
$g_{\sigma}$ is the coupling of the scalar meson to the free nucleon, which is, of course, directly related to the coupling of the scalar field to the quarks. The scalar polarizability, $d$, which quantifies the effects of the scalar field on the nucleon structure, is determined within the model and is well approximated as $d\approx$0.18 $R_B$, with $R_B$ being the bag radius (set to $1$ fm, 
see Ref.~\cite{GUICHON2018262}). The coupling of the nucleon to the vector fields, $g_{\omega}\omega$ and $g_{\rho} \rho$, does not affect the internal structure of the bag but rather contributes a constant shift to its energy. With increasing density, the reduction in strength of the $\sigma$ coupling to the nucleon is of key importance and forms the basis for the saturation mechanism of nuclear matter. Another crucial aspect of 
the QMC model is that the change in hadron structure in-medium  
provides a natural mechanism to generate three-body forces between all 
hadrons~\cite{Guichon:2004xg}, without additional parameters.

A comprehensive summary of the model can be found in the recent review \cite{GUICHON2018262}, which includes an account of the development of 
an energy density functional (EDF), derived from the underlying relativistic quark model~\cite{Guichon:2006er}.  The first version of the EDF systematically applied to nuclear structure was QMC-I~\cite{Stone2016PRL}, followed by QMC$\pi$-I \cite{Stone2019}, which included the contribution of a long-range Yukawa single pion exchange, and QMC$\pi$-II \cite{Martinez2019}, which took into account the non-linear self-interaction of the $\sigma$-meson. This generalization allows a contribution of the $\sigma$ exchange in the \textit{t-channel} to the polarizability, that cannot arise from the response of the bag \cite{GUICHON2018262}. It involves an additional parameter $\lambda_3$ which has to be obtained from a fit to experiment but leads to a significant improvement within QMC$\pi$-II of the predictions of the saturation properties of symmetric nuclear matter \cite{Martinez2019}.

In this work we use the latest version of the model, QMC$\pi$-III~\cite{Martinez2020}, 
which retains all of the features of QMC$\pi$-II and 
in addition includes the previously neglected spin-tensor $\vec{J}$ 
terms, which arise naturally within the QMC model. These additional terms 
do not add any more adjustable parameters, as their coefficients are 
calculated within the model. Thus, the QMC$\pi$-III model depends in 
total on only five free parameters: the $\sigma$ meson mass,  
$m_{\sigma}$, the effective meson coupling constants, $G_{m}= g^2_m / 
m^2_{m}$ (where $m$ stands for different mesons $m = \sigma, \omega, 
\rho$), and the $\sigma$ self-interaction parameter, $\lambda_3$. The 
parameters are fitted to 162 data points, consisting of the binding energies 
and root mean-square charge radii of seventy semi- and doubly- magic 
nuclei. The remaining model inputs are the $\omega$- and $\rho$- meson 
masses and the isoscalar and isovector magnetic moments (which appear in 
the spin-orbit interaction), taken at their physical values.

In the standard formulation of HF+BCS mean-field models, the EDF is 
augmented with pairing and Coulomb contributions. In the previous 
versions of the QMC model the volume pairing was used, with the pairing 
amplitudes calculated using BCS theory and the parameters of the pairing 
force being treated as additional fitting parameters. In the 
QMC$\pi$-III model we employ a density-dependent delta interaction where 
the usual pairing strengths are fully expressed in terms of the QMC 
model parameters. Both the direct and exchange terms of the Coulomb 
interaction were included as in the previous QMC calculations.

The final parameters were determined as $G_\sigma = 9.619$ fm$^2$, 
$G_\omega = 5.213$ fm$^2$ and $G_\rho = 4.712$ fm$^2$ couplings, the 
$\sigma$ self-interaction $\lambda_3 = 0.048$ fm$^2$ and the mass of the 
$\sigma$ meson $m_\sigma = 506$ MeV.

The QMC EDF has been implemented in the computer code SKYAX, to 
calculate ground-state properties of even-even axially symmetric and 
reflection-asymmetric nuclei. The code, which was originally designed to 
use Skyrme-type EDFs, has been adapted by P.-G. Reinhard 
\cite{Stone2016PRL, PGReinhard} and further modified by Martinez \textit{et 
al.}~\cite{Martinez2020} to include new features of the QMC$\pi$-III 
model. Experimentally unknown masses of nuclei on the neutron rich side 
of the nuclear chart were calculated and used in this work, in the region 
defined approximately by $20 < Z < 50$ and $50 < N < 90$.

We note that versions of the QMC model, differing from the formulation 
adopted in this work in that they do not include self-consistent 
solution of the field equations, have also been applied to model NSs and 
dense matter. They are listed and briefly discussed in 
Ref.~\cite{GUICHON2018262}.

\subsection{Modelling of the neutron star outer crust}\label{sec:OCcalc} 

We are calculating the outer crust that is at densities below the 
neutron drip line, and are taking the temperature of the NS as being 
$\sim 10^6$K, which is below the crystallization temperature of 
iron. We assume the simple scenario that the layers of the outer crust 
are one-component pure bcc Coulomb crystals of fully ionized atoms and 
the electrons are fully degenerate and can be treated as a uniform ideal 
relativistic Fermi gas.

The equilibrium condition of the system is determined by the 
minimization of the Gibbs free energy per nucleon, $g$, defined as 
\cite{PhysRevC.83.065810}.
\begin{equation}\label{eq:Gibbs} 
g = e + \frac{P}{n} \, , 
\end{equation} 
where $e$ is the energy per nucleon, $P$ is the total pressure at the 
given point in the NS crust and $n$ is the nuclear matter density,  
calculated numerically for a given pressure.

The energy per nucleon in the outer crust is defined as
\begin{equation}\label{eq:Mprime}
e = M'(A,Z)/A + \mathcal{E}_e/n + E_L(A,Z)/A \, ,
\end{equation}
having three terms representing contributions from nuclei, free 
electrons and lattice structure. In the first term, $M'(A,Z)$ is the 
atomic mass of the element (A, Z) with the binding energy of the atomic 
electrons subtracted out
\begin{equation}
    M'(A,Z) = M(A,Z) + (1.44381\cdot10^{-5} Z^{2.39} + 1.55468\cdot10^{-12} Z^{5.35}).
\end{equation}
As the only nuclear physics input, the atomic masses $M(A,Z)$ are 
tabulated in Wang and Audi (2017) \cite{Wang_2017} for experimentally 
available nuclei and are otherwise calculated with different theoretical 
models. The second term of Eq.(\ref{eq:Mprime}) contains the electron 
energy density
\begin{equation}\label{eq:ue(ne)}
\mathcal{E}_e(n_e) = \mathcal{E}^0_e(n_e) (1.00116 - 1.78 
\cdot10^{-5}Z^{4/3}),
\end{equation} 
where $\mathcal{E}^0_e(n_e)$ is the energy density of a uniform free 
electron gas with number density $n_e = Z n/A$ at $T=0$, as in 
Eq.(24.158) of Ref.\cite{2004cgps.book.....W}. The second factor of $\mathcal{E}(n_e)$ has two terms, the first assessing the electron 
exchange term and the second representing deviations of the electron gas 
from uniformity \cite{Haensel1994}. Finally, the last term of 
Eq.(\ref{eq:Mprime}) comes from the energetically most favourable bcc 
lattice configuration of ions, with lattice energy
\begin{equation}
E_L (A,Z) = -0.89593 \frac{Z^2 e^2}{R}
\end{equation}
where point nuclei are assumed, defining the ion radius by 
$\frac{4}{3}\pi R^3 =\frac{A}{n}$ and not taking into account the finite 
size nuclear correction.

The total pressure $P$ at any point of the outer crust has only electron 
and lattice contributions
\begin{equation}\label{eq:P}
P = P_e + P_L \, ,
\end{equation}
since nuclei do not exert any pressure at $T=0$. The electron pressure, 
taken from \cite{chandrasekhar1939introduction}, is
\begin{equation}\label{eq:Pe(ne)}
    P_e = P_e^0(1.00116 - 1.78\cdot 10^{-5} Z^{4/3}),
\end{equation}
with $P_e^0$ taken from Eq. (24.158) of Ref. \cite{2004cgps.book.....W} 
and the second factor contributions being the same as for the electron 
energy density (see appendix \ref{sec:ApendixB} for details). The 
lattice pressure is
\begin{equation}
P_L = \frac{n}{3}\frac{E_L}{A}.
\end{equation}
The nuclear matter density n is found numerically through Eq. (\ref{eq:P}) as the density that returns the total value of the pressure P.

The atomic mass tables are searched through in order to find nuclei that 
minimize $g$ for a certain value of pressure and the corresponding density. 
The assumption is that for each value of $P$ only one nucleus would 
appear, resulting in density discontinuities between consecutive nuclear  
species. These are assumed to be coexistence regions of neighbouring 
nuclei and are not considered further.

As by definition there are no free neutrons in the outer crust, the 
outer-inner crust boundary is defined by the condition
\begin{equation}\label{eq:equilibrium}
    \mu_n \leq 0,
\end{equation}
with $\mu_n$ being the neutron chemical potential. As long as Eq.(\ref{eq:equilibrium}) holds, all the neutrons are bound in nuclei forming a lattice. In a NS environment, the neutron chemical potential is calculated through the identity
\begin{equation}\label{eq:mu_n}
    \mu_n = g - m_n
\end{equation}
which is valid because of the beta equilibrium holding in the NS (see Ref. \cite{Baym:1971pw} and appendix A of Ref. \cite{Pearson2018}). As 
the density increases, the value of $\mu_n$ rises monotonically and eventually becomes positive, meaning that the neutron drip line has been 
reached and that the last nuclide in the outer crust of the NS has been identified.

The importance of further corrections, such as those for screening of 
electrons by protons (and protons by electrons), for the 
electron-correlation energy, for zero-point energy, finite nuclear size 
corrections and thermal corrections has been addressed in more detail in 
Ref.~\cite{PhysRevC.83.065810} and they were mostly found to play a 
negligible role. Pearson {\em et al.}~\cite{Pearson2018} did include these 
corrections and used complete expressions for the electron exchange and 
screening corrections to the electron energy density (the screening 
sometimes being referred to as ‘polarization’). A significant influence 
of the polarization correction was found in the determination of the 
crystallization temperature, at which the plasma ocean at the NS surface 
would crystallize and settle into a lattice structure, marking the 
beginning of the solid outer crust. However, as the crystallization 
temperature is expected to vary between $\sim 10^8$ K - $10^9$ K and the 
temperature assumed in this work is $\sim 10^6$ K, these corrections 
have not been included here.

Theoretical nuclear masses are calculated in the QMC$\pi$-III model and 
experimental data are taken from the 2016 Atomic Mass Evaluation 
\cite{Wang2017}. We note that both directly measured and indirectly 
determined masses~\cite{Huang2017} were taken as experimental.

\subsection{Nuclear models for neutron rich nuclei}\label{sec:models}

It is interesting to compare the composition of the outer crust of the 
NS as given by the QMC$\pi$-III model with those given by other 
theories. This can easily be done by replacing the QMC$\pi$-III masses 
in our formalism with those given by the other models and looking for 
the differences. In all cases we compare only results for even-even 
nuclei, as the QMC model for odd-A and odd-odd nuclei has not yet been developed. 
The nuclear chart region of interest for this study is fairly 
restricted, spanning the proton numbers $26 < Z <50$ and 
the neutron numbers $ 30 < N < 90$, since only nuclei in this region are 
expected to appear in the NS outer crust, as will be discussed in 
Sec.~\ref{sec:Results}.

We use the finite range droplet model FRDM~\cite{Moeller2012a,MOLLER20161} as an example of the liquid-drop based macroscopic-microscopic models. This model provides predictions of the atomic mass excesses and binding energies, ground-state shell-plus-pairing corrections, ground-state microscopic corrections, and nuclear ground-state deformations of 9318 nuclei in the range from $^{16}$O to A=339.  It depends on 17 constants adjusted to nuclear masses or mass-like quantities, 21 determined from other considerations (including 5 fundamental constants) and a number of empirical relations detailed in Ref.~\cite{MOLLER20161}.

As an example of a mass model based on a non-relativistic EDF with the Skyrme interaction, we use a member of the Brussels-Montreal family of 
models. The latest models of this family used to explore the NS outer crust properties were of series HFB-22 to HFB-26~\cite{Goriely2013, 
Chamel2016,Pearson2018}. The HFB-24 model with the BSk-24 Skyrme interaction produced the best fit to the 2012 AME database of 2353 nuclear masses, as well as properties of nuclear matter and neutron  
stars~\cite{Goriely2013,Chamel2016}. The nuclear mass excess data is available via the Brussels Nuclear Library for Astrophysics Application (BRUSLIB)~\cite{BRUSLIB}. The Skyrme energy density functional is dependent on 16 variable parameters. In addition, the specific pairing force has 5 
parameters, the Wigner term, needed to improve fits to N=Z masses, depends on 4 parameters and the correction for spurious collective motion has 5 parameters. In total, the HFB24 model is determined by 30 adjustable parameters.

Masses of even-even nuclei calculated in the RMF model based on the Lagrangian density of the Walecka model with the NL3 parameterization 
were also used in this work. That model utilizes the exchange of $\sigma, \omega$ and $\rho$ mesons between nucleons in the mean-field and no-sea approximation \cite{LALAZISSIS19991,Lalazissis1997} and has the quadratic scalar potential in the Lagrangian replaced by a quartic form including non-linear $\sigma$ self-interaction terms. The model has 6 parameters: the mass of the $\sigma$ meson, 3 meson-nucleon coupling constants and 2 for the $\sigma$ meson self-interaction.

It is interesting to compare predictions of the models for basic properties of symmetric nuclear matter at the saturation density. Of particular interest is the symmetry energy, the difference between the energy per particle in pure neutron matter and in symmetric matter (with equal numbers of protons and neutrons) expressed as the symmetry energy coefficient $J$ in the semi-empirical mass formula, its slope $L$ and the incompressibility $K$.

The QMC$\pi$-III parameterization yields symmetric nuclear matter properties at saturation: the saturation density $\rho_0$ and the energy per particle E$_0$, the asymmetry coefficient a$_{sym}$, the incompressibility $K$ and its slope $L$, to be $\rho_0 =0.15$ fm$^{-3}$, 
E$_0 =-15.7$ MeV, a$_{sym} = 29.42$ MeV, K$= 233$ MeV and L$ = 43$ MeV. The asymmetry parameter a$_{sym}$ is closely related to the more widely used symmetry energy coefficient $J$ and for practical purposes the two 
quantities may be taken to be equal (for a discussion see Ref.~\cite{Stone2007}, 
Sec. 4.2.1).  We summarize these quantities for QMC$\pi$-III, HFB-24, FRDM and NL3 in Table~\ref{tab:sym}.

\section{Results and discussion}\label{sec:Results}

\subsection{QMC$\pi$-III nuclear mass table}

\begin{figure*}[t]
 	\centering
 	\includegraphics[angle=0,scale=0.5]{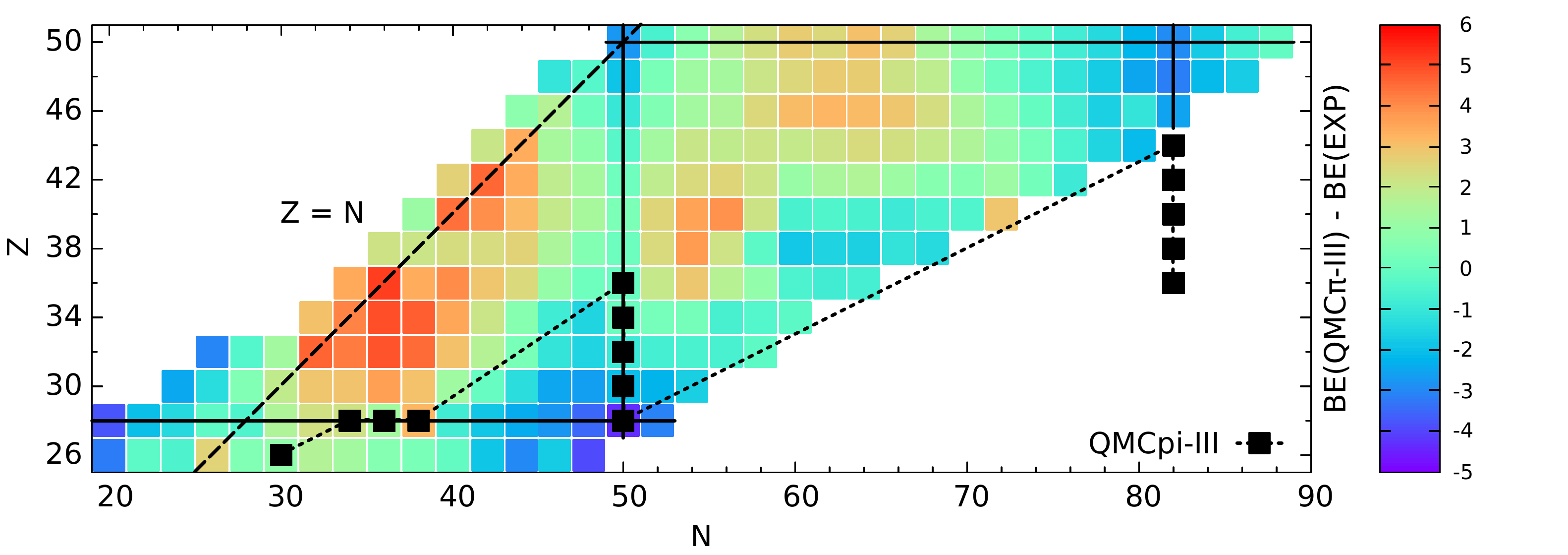}	
 	\caption{Differences between nuclear 
 	binding energies (BE) as measured experimentally and as calculated with the QMC$\pi$-III model: the color code shows the magnitude of the difference for each nucleus, measured in units of MeV. The comparison is given just for the region of the chart which is of interest for the outer crusts of neutron stars. The magic numbers $Z=28, 50$ and $N=50, 82$ are indicated with full lines while the dashed line corresponds to $Z=N$. The nuclei marked in black are those involved in building the outer crust, according to the QMC$\pi$-III mass model. \label{fig:EXPvsQMC}}
 \end{figure*}

In order to demonstrate the precision of the QMC$\pi$-III~\cite{Martinez2020} model for calculating nuclear binding energies, Fig.\ref{fig:EXPvsQMC} shows, using a color code, the differences between the calculated and experimentally measured values for those nuclei where measured ones are available. The sequence of nuclei 
building the NS outer crust is shown with the black squares in the figure. Up to $^{78}Ni$, the experimental values are used directly in the further calculations (since they are available) while for heavier nuclei (starting with $^{126}Ru$ for QMC$\pi$-III), values calculated from the model are used. The nuclei appearing in the outer crust sequence are mainly semi-magic, having magic numbers of either neutrons N or protons Z, indicating the role of stabilising shell effects. The present version of the QMC$\pi$-III model tends to overestimate 
the binding energies for nuclei with (semi-)closed shells in comparison 
with the HFB24 and FRDM models, which give binding energies closer to the 
experimental ones in some cases, but at the price of having a larger number of model parameters. The origin of this feature of the QMC model is so far unknown and will be addressed in further developments of the model.

\subsection{Equation of state (EOS)} \label{sec:EOS}

\begin{figure*}[b]
	\centering
	\includegraphics[scale=0.55]{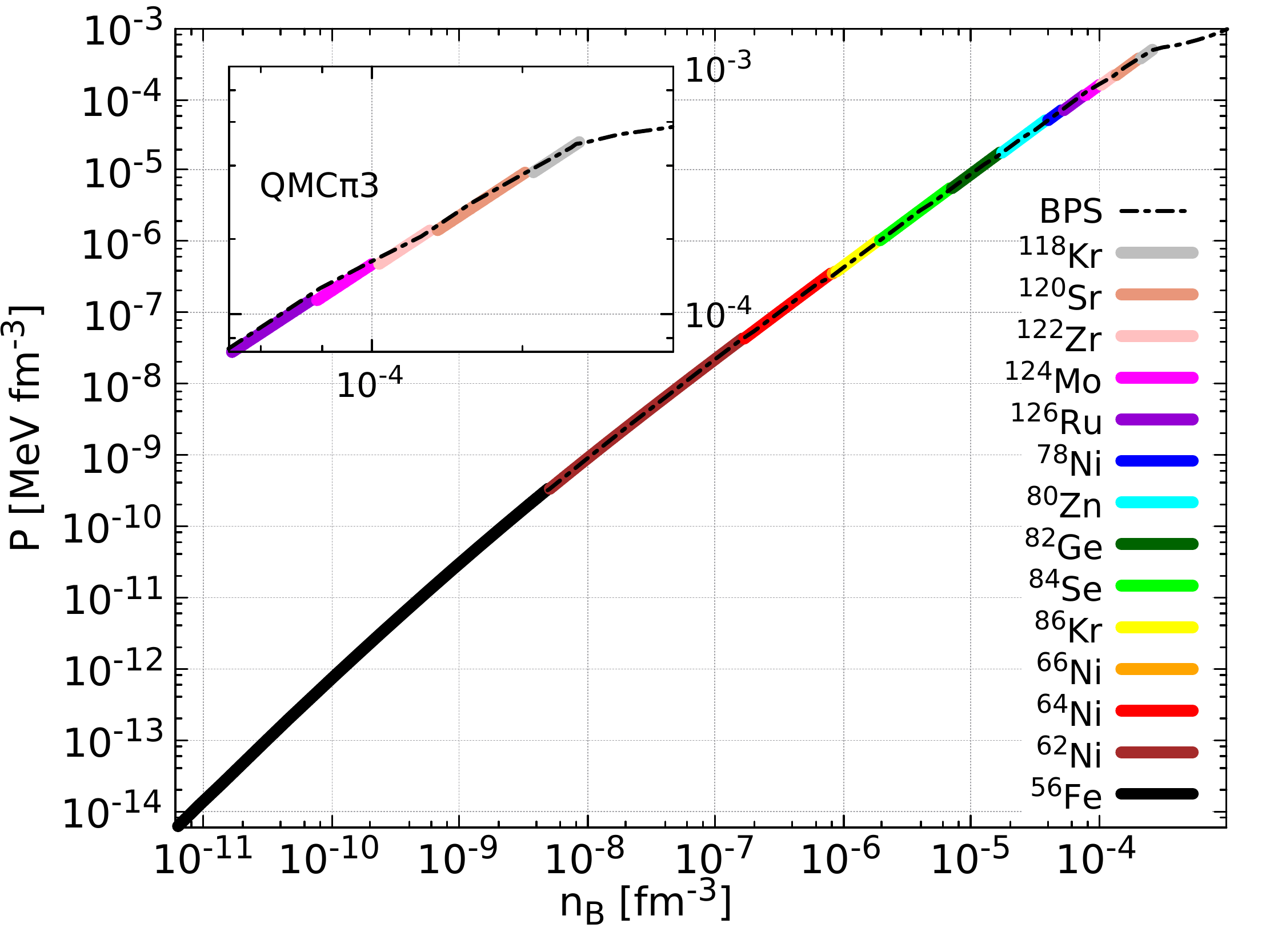}
	\caption{The EOS for the ground state of the outer crust as given by the QMC$\pi$-III model. The colour code indicates the different ion species involved. The BPS EOS is shown for comparison, marked with the dashed line.}
	\label{fig:EOS}
\end{figure*}

The equation of state (EOS) for matter in the outer crust
of the NS is calculated starting from the NS surface and going inward through the crystal lattice of different nuclei until the boundary with the inner crust is reached. For each value of pressure ($P=0$ at the surface) the Gibbs free energy (Eq. \ref{eq:Gibbs}) is minimized and the nuclei building the crust are determined. The EOS for the NS outer crust calculated within the QMC$\pi$-III model is illustrated in Fig.~\ref{fig:EOS} where various nuclear species are presented in different colors. At the NS surface, the $^{56}$Fe nuclei are energetically favored up to energy density $4.568\cdot 10^{-6}$ MeV fm$^{-3}$, followed by the nickel Ni isotopes $^{62,64,66}$Ni and heavier nuclei with closed neutron number shell at $N=50$, $^{84}$Kr, $^{82}$Ge, $^{80}$Zn and $^{78}$Ni. Up to this point the sequence of nuclei is well established, since it depends only on the experimental masses. Beyond an energy density of around $4.726\cdot 10^{-2}$ MeV fm$^{-3}$, the masses of the following nuclei in the sequence, $^{126}$Ru, $^{124}$Mo, $^{122}$Zr, $^{120}$Sr and $^{118}$Kr, with closed neutron shell $N=82$, are obtained theoretically within the QMC model
(shown in the inserted plot of Fig.~\ref{fig:EOS}). The line is not continuous since we have excluded coexistence regions between two neighbouring nuclei. 
The pressure increases in steps of $\delta$P $= 0.003$ P until the maximum density of n$_B^{max} = 2.61 \cdot 10^{-4}$ is reached, beyond which we enter the inner crust of the NS where free neutrons appear alongside free electrons.

For comparison, the extensively used Baym-Pethick-Sutherland (BPS) outer crust  
EOS~\cite{Baym:1971pw} is also included in 
Fig.~\ref{fig:EOS}. The EOSs of the other models from section~\ref{sec:models} are not plotted because the differences between them are not distinguishable within the resolution of the plot. This is not surprising since the pressure of the matter is dominated by the electron contribution, $P_e$, which is the same in all of the models.
 
\subsection{The sequence of nuclei and the two-neutron drip line}

The differences between the theoretical models do not have a significant influence on the EOS but the sequence of nuclei building the NS outer crust is model dependent. The numerical results of the EOSs are presented in Table~\ref{tab:EOS_QMCIII}, listing the minimum and maximum baryon number densities (n$_{min}$ and n$_{max}$) for the layers corresponding to the different nuclear species, as well as the values of the pressure P and energy density $\epsilon$ at the boundaries between them. The last two columns give the neutron and electron chemical potential values at the boundaries of each element layer. The nuclear masses in the upper part of the table are taken from the experimental mass table of Wang and Audi (2017)~\cite{Wang_2017}, while 
those in the lower parts of the table were calculated 
with the various theoretical models under consideration. The $^{78}$Ni nucleus is included in both sections since its mass is experimentally known but the value of $n_{max}$ for the $^{78}$Ni layer is model dependent.

\begin{figure*}[t]
	\centering
	\includegraphics[scale=0.65]{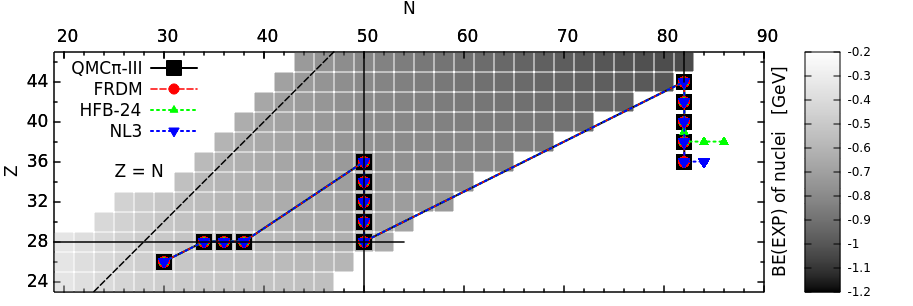}\\
	\includegraphics[scale=0.65]{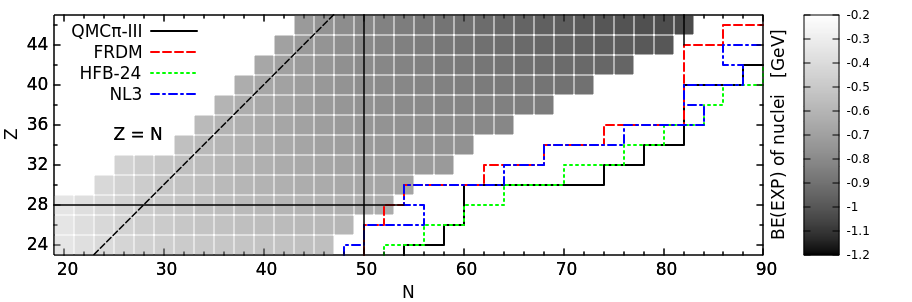}
	\caption{\label{fig:Nsequence} Top: The sequence of nuclei involved in building the outer crusts of neutron stars, as predicted by the QMC$\pi$-III, FRDM, HFB24 and NL3 mass models. Bottom: The two-neutron drip line for the same set of mass models. The grey-scale indicates the experimentally determined binding energies of the nuclei in GeV, where available. The full black lines indicate N and Z magic numbers while the black dashed line corresponds to $Z = N$.}
\end{figure*}

The sequences of outer crust nuclei for the different theoretical models are indicated in the top panel of Fig.~\ref{fig:Nsequence} by different colors. All of these models predict nuclei along the $N=82$ line with variations in the deepest layers of the outer crust, before the neutron drip-line (the boundary between the inner and outer crust) is reached. As the current version of the HF+BCS method used in this work is designed to calculate only properties of even-even nuclei, we can determine the transition to the inner crust only as a two-neutron drip-line, whose exact position is model dependent, as demonstrated in the bottom panel of Fig.~\ref{fig:Nsequence}. All of the models being used here predict this transition to happen for density around $\sim 2.6\cdot 10^{-4}$ fm$^{-3}$.
 
\subsection{TOV equations for the neutron star outer crust}
\label{sec:TOV}

Details of the structure of the outer crust can be 
obtained by integrating the well-known Tolman-Oppenheimer-Volkoff (TOV) equations for hydrostatic equilibrium~\cite{PhysRev.55.364, PhysRev.55.374},
\begin{equation} 
\frac{dP(r)}{dr} = -\frac{G \rho(r) 
\mathcal{M}(r)}{r^2} \Big[1+\frac{P(r)}{c^2 \rho(r)}\Big] 
\Big[1+\frac{4\pi P(r)r^3}{c^2\mathcal{M}(r)}\Big] 
\Big[1-\frac{2G\mathcal{M}(r)}{c^2 r}\Big]^{-1} 
\end{equation} 
 and 
\begin{equation} 
\mathcal{M}(r)=4\pi \int_0^r \rho(r')r'^2 dr' \, , 
\end{equation} 
where $\rho(r)$ is the mass density (at radius $r$) that appears in the EOS and $\mathcal{M}(r)$ is the (gravitational) mass internal to radius $r$. For doing this, it is necessary to have appropriate values for the mass and radius at the inner edge of the outer crust. To obtain those, we calculate the structure of the inner parts of the NS, starting from the center and using a composite EOS for the whole star comprising (i) a QMC$\pi$-III based part for the central core, including the full baryon octet~\cite{GUICHON2018262}, (ii) the BPS EOS covering the inner crust (since a QMC$\pi$-III model is not yet available for that region) and  (iii) the QMC$\pi$-III EOS for the outer crust, as calculated in this work.
We first specify the required total gravitational mass of the star and iterate integrations of the TOV equations to find the corresponding central density. With that, we then make a further integration to find the mass and radius values at the inner edges of the inner and outer 
crust, iterating to determine those with high accuracy. The transition density between the outer core and the inner crust is taken to be where the QMC$\pi$-III high density EOS is joined to the BPS one (at $\sim 0.6n_0$), with the transition between the inner and outer crust coming at the neutron drip point, as mentioned earlier. This composite EOS gives a maximum NS mass of $1.97\,M_\odot$, with a radius of 
$12.28\,\rm{km}$, 
while for a $1.44\,M_\odot$ NS model it gives a radius of
$13.00\,\rm{km}$, which is compatible with the observational values recently obtained for PSR J0030+0451 using NICER \cite{Miller2019} ($1.44^{+0.15}_{-0.14}\,M_\odot$ with a radius of $13.02^{+1.24}_{-1.06}\,\rm{km}$ at $68\%$ confidence level). In Table~\ref{tab:comp} we show results obtained with the composite EOS for different parts of the star (core, inner crust and outer crust) for NS models with masses of 1.0, 1.4 and $1.94\,M_\odot$ within the QMC$\pi-III$ model. Our other theoretical models (FRDM, NL3 and HFB-24) give very similar values to the QMC$\pi$-III ones and are therefore not included in the table.

\begin{figure}[t]
        \centering
        \includegraphics[scale=0.6]{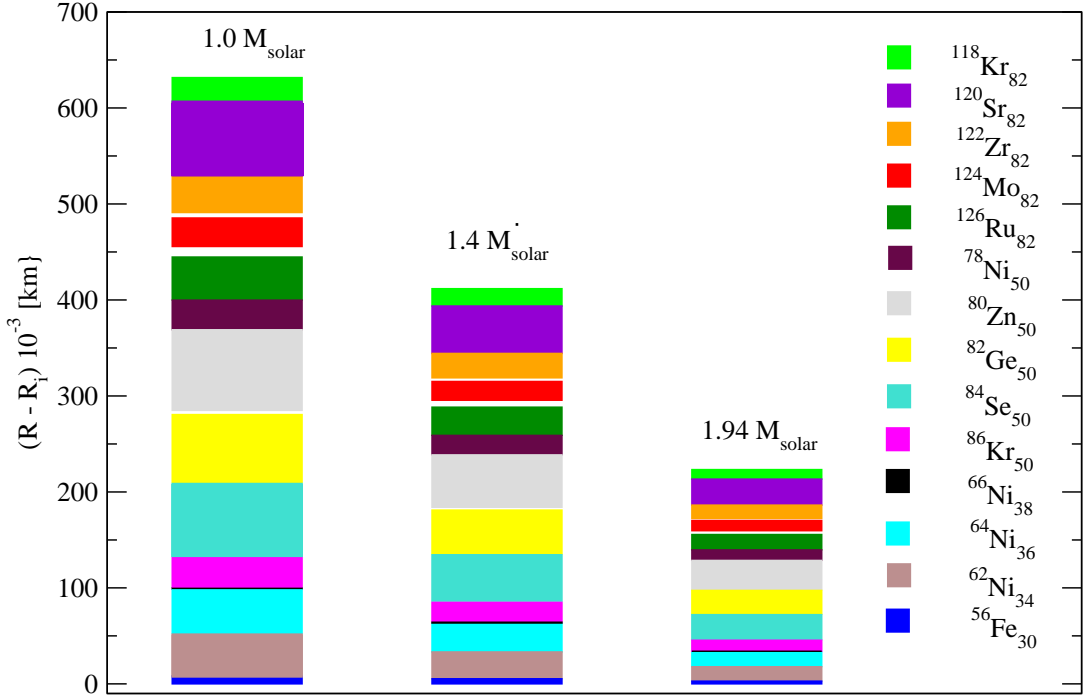}
        \caption{Illustration of the sequence of single-ion pure layers in the outer crust of three neutron star models with gravitational masses of 1.0, 1.4 and $1.94\,M_\odot$, as calculated using the QMC$\pi$-III model: $R$ signifies the total radius of the star, and $R_{i}$ labels the radii of the top and bottom of each layer. The white spaces between some layers are the transition regions (gaps) containing more than one species of ion. The gaps have not been included in the present analysis.}
\label{fig:stack}
\end{figure}

For obtaining the depths and masses of the successive layers in the outer crust, corresponding to the succession of nuclei listed in Table~\ref{tab:EOS_QMCIII} for QMC$\pi$-III, we then make a high-resolution integration of the TOV equations in this region, for NS models with each specified gravitational mass. For each layer, we calculate the radii corresponding to the maximum and minimum densities in the layer and the gravitational mass contained within it. 
Table~\ref{tab:layers} shows results for the same NS models as in Table~\ref{tab:comp} with the mass in each layer given as a percentage of the total mass in the outer crust. Table~\ref{tab:layers} also contains results from similar calculations carried out for the three alternative comparison EOSs. 
The results from the QMC$\pi$-III outer crust calculation are also illustrated in Fig.~\ref{fig:stack}, with the profiles being shown as a function of the distance from the surface of the star. It is interesting to observe the relation between the depth of the crust and the mass and radius of the star.

\subsection{Adiabatic index}

\begin{figure}[t]
	\centering
	\includegraphics[scale=0.5]{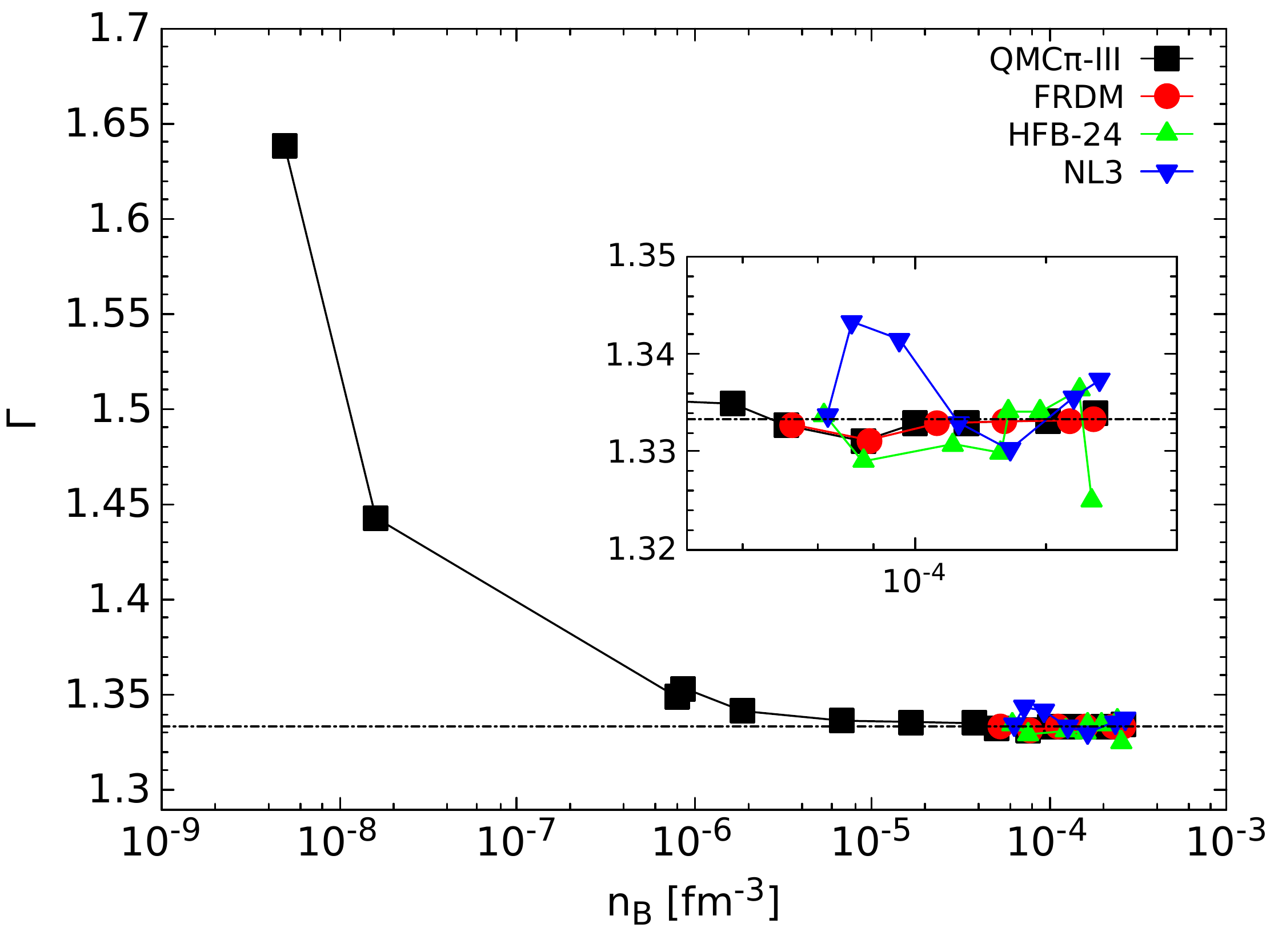}
	\caption{The adiabatic index $\Gamma$ of the ground state of the outer crust as calculated using the QMC$\pi$3, FRDM, HFB24 and NL3 mass models, plotted as a function of the baryon number density n$_B$. The plot labelled QMC$\pi$3 includes both results obtained using experimental masses, where available, and also ones using theoretical masses calculated with QMC$\pi$3. For the other mass models only results obtained with theoretical masses are shown (see the insert for details). The horizontal dashed line corresponds to $\Gamma$=4/3.}
\label{fig:gamma}
\end{figure}

An important property of an EOS is its stiffness which 
can be represented by the dimensionless adiabatic index, defined as~\cite{Chamel2008}
\begin{equation}
\Gamma = \frac{d \log P}{d \log n_b}=\frac{n_b}{P}\frac{d P}{d n_b} \, ,
\end{equation}
 with $P$ being the total pressure and $n_B$ the baryon density. At subnuclear densities, $\Gamma$ can be approximated as $\frac{\epsilon}{P} \frac{d P}{d\epsilon}$ using the energy density $\epsilon = \rho c^2$ \cite{Chamel2008}. The outer crust pressure in the ground state is well approximated by Eq.~\ref{eq:P}. Figure~\ref{fig:gamma} shows the adiabatic index as a function of baryon number density for all of the layers in the outer crust as calculated using the QMC$\pi$-III model (with the gaps excluded). As can be seen, at high mass densities $\Gamma$ approaches the limit of 4/3. This is because both the electron and lattice pressures, $P_e$ and $P_L$, have energy density dependence proportional to $\sim \rho^{(4/3)}$. It is instructive to compare this result with Fig.~3 in 
Ref.~\cite{PhysRevC.73.035804} which shows a very similar behaviour for a large range of models. This seems to suggest that the stiffness of the outer crust is rather insensitive to the choice of different mass models.

\subsection{Speed of sound}

\begin{figure}[t]
	\centering
	\includegraphics[scale=0.5]{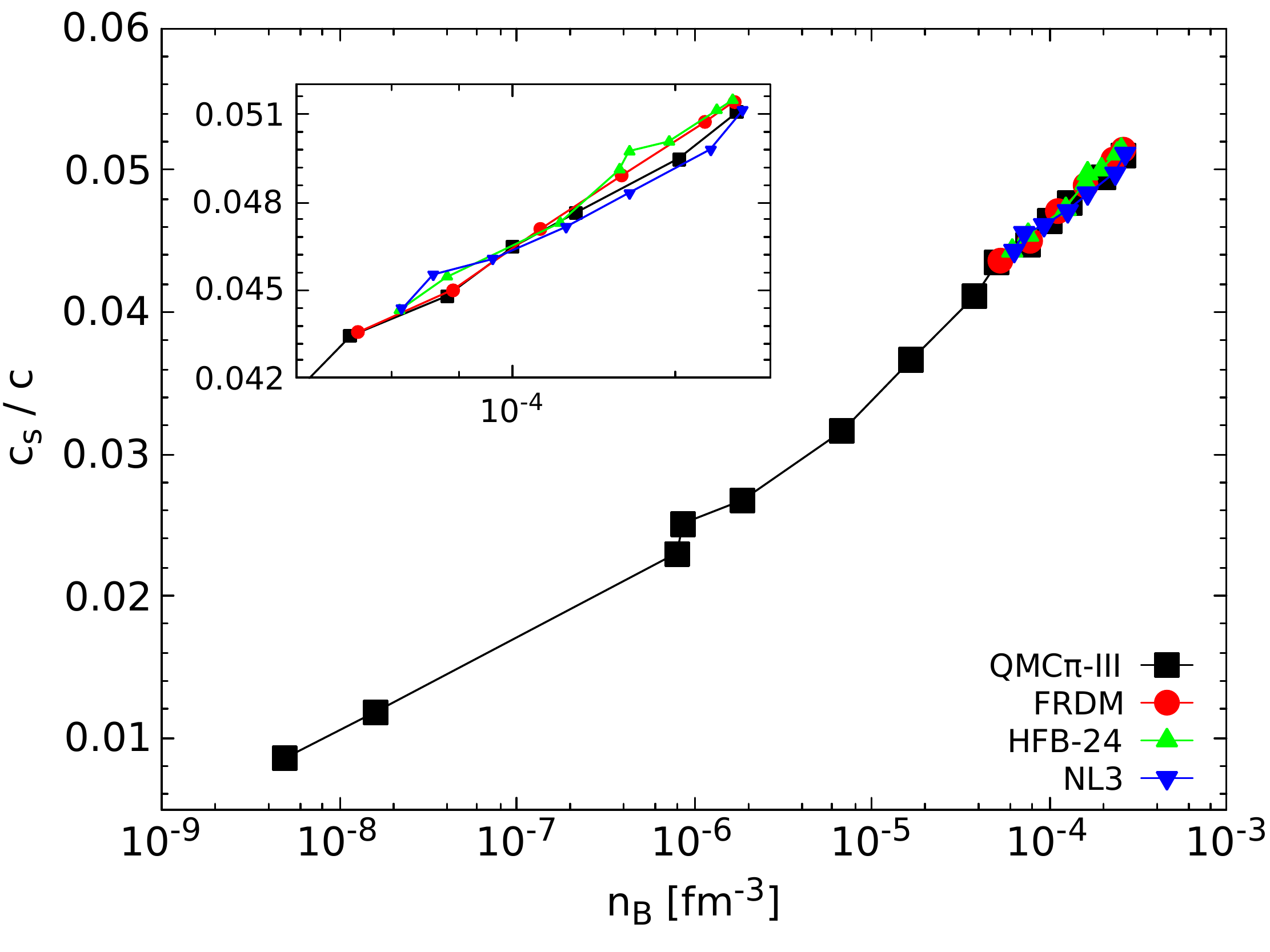}
	\caption{Similar to Fig.~\ref{fig:gamma} but for the speed of sound c$_S$ in units of the speed of light $c$.}
\label{fig:csound}
\end{figure}

In Figure~\ref{fig:csound} we show the behavior of the 
speed of sound defined as
\begin{equation}
\Big(\frac{c_s}{c}\Big)^2=\frac{dP}{d\epsilon}
\end{equation}
in units of the speed of light $c$. Although the speed of 
sound in the outer crust is very low (as expected), it is interesting to observe that it  grows systematically with increasing particle number density of the medium.
This is also consistent with predictions of the speed of sound in NS matter based on chiral effective field theories and the AV8$^\prime$+UIX interaction~\cite{Tews2018}. We observe that all four mass models used in this work give very similar answers for this, showing a limited sensitivity of the speed of sound to the composition of the outer crust.

\subsection{Specific heat of the outer crust}

The specific heat of baryonic matter in the inner crust of a neutron star provides the microscopic input for the solution of the heat transport equations, governing thermalization of neutron stars (see e.g. Ref.~\cite{Fortin2010}). The cooling rate is dependent on the ratio of the specific heat to the thermal conductivity and proportional to the square of the crust depth. Therefore the outer crust, only several hundred meters deep, plays a minor, but non-negligible role in NS cooling. Thermal properties of the outer crust are fully determined by the electrons. For completeness, we calculate the electron specific heat capacity C$_e$ for each layer, 
assuming that the electrons are a nearly ideal, strongly degenerate, ultra-relativistic gas. At constant temperature and pressure~\cite{Yakovlev1999},
\begin{equation}
C_e \sim 3.54\times10^{-14}\Big(\frac{n_e}{n_0}\Big)^{2/3} T_9 
\hspace{2mm} \hspace{0.5mm} \textrm{[MeV fm$^{-3}$ K$^{-1}$]} \, ,
\end{equation}
 where $n_0$ is the saturation density and $T_9$ stands for $T_9 = T/10^9$ K. The results for the QMC$\pi$-III model of a NS with gravitational mass $M_g =1.4\,M_\odot$ and temperature $T = 10^6$ are summarized in Table~\ref{tab:C_e}, together with the ion and electron number densities and the Fermi momentum of the electrons.

\section{Summary and conclusions} \label{sec:Conclusion} 

In this paper we have calculated zero-temperature ground 
state properties of the outer crust of non-accreting, non-rotating neutron stars using measured masses for the nuclei where available and otherwise, for the first time, theoretical predictions coming from the QMC$\pi$-III model. The FRDM, HFB24 and NL3 mass models have been used as alternatives for comparison with QMC$\pi$-III. The EOS for the outer crust was found to be substantially the same for all four models but the nuclidic composition of the layers 
closer to the neutron drip line, and the drip line itself, varied from model to model. We found the last nuclide before the drip line to be $^{118}$Kr in the QMC$\pi$-III and FRDM models, $^{124}$Sr in the HFB24 model (in agreement with~\cite{Chamel2016}), and $^{120}$Kr in the NL3 model. 

The structure of the outer crust was found by integrating the TOV equations with the above EOSs. Of course, in order to obtain suitable initial conditions at the inner edge of it, we needed to also integrate over the inner regions of the NS. For that we used QMC$\pi$-III for the core and the BPS EOS for the inner crust. Once the mass and radius values were determined at the inner edge of the outer crust, the continuing outward TOV integration then gave results for the depths and masses of the succession of layers comprising it (each characterised by a different nuclide). NS models with $1.0$, $1.4$ and $1.94\,M_\odot$ were studied. For each case, we observed only slight variations in the results when varying the EOS used. As all models are fitted to the same set of data when the data is known, the close agreement between them is not surprising. The extrapolation to unknown masses in this work is not going sufficiently far away from experiment for the model differences to be exposed. Their reliability, however, can be judged by the number of unknown parameters. With only five, well-controlled parameters related to underlying background physics, QMC has an advantage over the more traditional, multiple parameter models and stands a good chance 
of providing reasonable predictions for the experimentally inaccessible regions of the nuclear chart. Finally, we calculated the adiabatic index, the speed of sound, the specific heat, and the electron number density and Fermi momentum to complete the investigation of the performance of the QMC$\pi$-III model. Comparing these results with the literature, no anomalies were found. 

As for the future: in order to obtain the complete 
QMC EOS for neutron stars from crust to core, the inner 
crust still remains to be modeled. Ideally this would involve an extension to finite temperature. The temperature of neutron stars can be measured only at the surface and its variation going towards the center is model dependent. Significant changes in the composition and mechanical properties of the inner and outer crust have been suggested to occur during the cooling process after the birth of a neutron star (see e.g.  Refs.~\cite{Chamel2008,Potekhin2010,Yakovlev2010,Potekhin2015a}). An interesting connection between ejection of material from the neutron star crust and the r-process was investigated by Goriely {\em et al.}~\cite{Goriely2011}. They studied the composition of the outer crust material after the decompression that would follow a possible ejection and found that, depending on the initial state and temperature of the matter, the decompression could provide suitable conditions for a robust r-processing of the light species, particularly r-nuclei with A $<$ 140.

Furthermore, the crystalline structure of the crust, including the outer part, and its deformation and strength are not only temperature dependent but are also strongly affected by magnetic fields~\cite{Baiko2018,Fattoyev2018}. 
Models of neutron star cooling considering the surface and/or atmosphere (for example, emission of x-rays from the carbon atmosphere of CAS A~\cite{Ho2009a}) assume an iron heat-blanketing envelope of the star~\cite{Yakovlev2010}, with uniform average thermal distribution over the surface. However, hot spots on the surface of the isolated pulsar PSRJ0030+0451, recently identified in the study of a variety of x-ray emission patterns obtained from the NICER 
mission~\cite{Miller2019,Riley2019}, suggest that the surface thermal distribution may be more complicated. We are not aware of any model which includes the effect of the more complex structure of the outer crust, studied in this work, on neutron star cooling.

An isolated neutron star born in a core-collapse 
supernova is likely to experience fallback of neutron-heavy 
r-process material created in the ejecta~\cite{FRYER2004,Wong2014,Hix2014}. 
The question of whether this material can remain on the surface of a proto-neutron star and reach equilibrium is open. Deposition of energetic particles on the surface of the star may ignite nuclear reactions, similar to those discussed in the case of accreting 
envelopes~\cite{Chugunov2018}, which would make the composition of the outer crust more complicated.

We conclude that studies of the outer crust of neutron 
stars, which are usually neglected in neutron star modelling, deserve further attention, particularly in the context of the effort to determine the best possible values for the radii of 
neutron stars with known masses.

\section*{Acknowledgments}
J.R.S. and P.A.M.G. acknowledge with pleasure support and hospitality of CSSM at the University of Adelaide during visits in the course of this work. This work was supported by the University of Adelaide and the Australian Research Council through the ARC Centre of Excellence in Particle Physics at the Terascale (CE110001004) and grants DP180100497 and DP150103101.

\appendix

\section{The electron contribution to the outer crust}\label{sec:ApendixB}

Both the energy per nucleon and the pressure have a term coming from the free electron gas contribution. The energy density of free electrons given by Eq.(\ref{eq:ue(ne)}) contains the energy density at $T=0$ of a free 
uniform electron gas, $\mathcal{E}^0_e(n_e)$, which is
\begin{equation}
\begin{split}
\mathcal{E}_e^0 
&= \frac{8\pi}{3 h^3} (p_F^3 \epsilon_F) - P_e^0,
\end{split}
\end{equation}
as given by Eq.(24.161) in Ref.\cite{2004cgps.book.....W}. 
The momentum and electron pressure 
are
 \begin{equation}
 p_F = (\hbar c)(3\pi^2 n_e)^{1/3}
 \end{equation}
 and
\begin{equation}\label{eq:Pe0}
P_e^0  = \frac{8\pi}{3 h^3}\int_{0}^{p_F}\frac{(p^4/m)dp}{\sqrt{1+(p/mc)^2}}.
\end{equation}
Following the Chandrasekhar approach 
\cite{chandrasekhar1939introduction} by introducing the substitutions
\begin{equation}
\begin{split}
\sinh{\theta}&=p/mc \\
\sinh{\theta_F} &= p_F/mc,
\end{split}
\end{equation}
letting $x=sinh{\theta_F} = p_F/mc$ and defining the 
functions $g(x)$ and $f(x)$ to be
\begin{equation}
\begin{split}
g(x) &= 8x^3(\sqrt(1+x^2-1)-f(x)) \\
f(x) &= x(x^2+1)^{1/2}(2x^2-3) + 3ln(x+\sqrt(1+x^2)),
\end{split}
\end{equation}
the internal kinetic energy of the electrons per unit 
volume V is given by
\begin{equation}
\begin{split}
\mathcal{E}_e^0  &= \frac{\pi m^4 c^5}{3 h^3}g(x) 
= 3.746 \cdot 10^{-11} g(x) \textrm{ MeV/fm$^3$},
\end{split}
\end{equation}
while the pressure is equal to 
\begin{equation}
\begin{split}
P_e^0 &= \frac{\pi m^4 c^5}{3 h^3} f(x) 
= 3.746\cdot 10^{-11} f(x) \textrm{ MeV/fm$^3$}.
\end{split}
\end{equation}

\bibliography{SAntic_OCpaper}

\clearpage
 \begin{table*}[t!] 
 	\centering
 	\setlength{\tabcolsep}{9pt}
 	\def\arraystretch{0.95}
 	\caption{Symmetric nuclear matter properties at saturation: the particle number density $n_0$, the energy per particle E/A, the symmetry energy $J$ and its slope $L$ and the incompressibility $K$ as given by the different theoretical mass models employed in this work. The data are taken from \cite{MOLLER20161,Goriely2013,Lalazissis1997}. $^\ast$ The value of the incompressibility quoted for the HFB24 model is strictly speaking just the volume part, K$_v$.\label{tab:sym}}
	\vspace{3mm}
 	\begin{tabular}{l c c c c c}
 		\hline
 		\hline
    Model           & $n_0$ & E/A & $J$ & $L$ & $K$ \\
                        & fm$^{-3}$ &  MeV  &  MeV & MeV & MeV \\
 \hline   
   QMC$\pi$-III  &   0.15    &      -15.7 &     29.42  &  43.0             &  233.0 \\
   FRDM             &   $-$     &       $-$   &     32.2    &  53.5             &   240.0 \\
   HFB-24          &   0.1578 & -16.048  &     30.0    & $^{\ast}$ 46.4 &  $^{\ast}$245.5\\
   NL3               & 0.148     & -16.299  &     37.4    & $-$               & 271.76\\
 \hline
 \hline
 	\end{tabular} 
\end{table*} 

\clearpage
\begin{table*}[t!] 
	\centering
	\bgroup
    \setlength{\tabcolsep}{5pt}
	\def\arraystretch{0.6}
	\caption{Composition and EOS of the outer crust for the QMC$\pi$-III, FRDM, HFB24 and NL3 mass models (experimental masses \cite{Wang_2017} are taken for the first eight nuclei).  Baryon number densities at the bottom and top of each layer n$_{max}$ and n$_{min}$ are also shown as well as the neutron and electron chemical potentials. \label{tab:EOS_QMCIII} }
	\vspace{3mm}
	\begin{tabular}{l l c c c c c c c}
		\hline
		\hline
		Z & N & $^{A}X$ & n$_{min}$ & n$_{max}$ & P$_{max}$ & $\epsilon_{max}$ & $\mu_n$ & $\mu_e$ \\
		&  &  & [fm$^{-3}$] &  [fm$^{-3}$] & [MeV fm$^{-3}$] & [MeV fm$^{-3}$] & [MeV]   & [MeV]   \\
		\hline
        $26$ &  $30$&   $^{56}$Fe  & $0$                 & $4.91\cdot10^{-9}$ &	$3.34\cdot10^{-10}$& $4.568\cdot10^{-6}$ & $-8.96$ & $0.8$  \\
        $28$ &  $34$&	$^{62}$Ni  & $5.07\cdot10^{-9}$  & $1.62\cdot10^{-7}$ &	$4.31\cdot10^{-8}$ & $1.510\cdot10^{-4}$ & $-8.25$ & $1.99$ \\
        $28$ &	$36$&	$^{64}$Ni  & $1.68\cdot10^{-7}$  & $7.96\cdot10^{-7}$ &	$3.53\cdot10^{-7}$ & $7.413\cdot10^{-4}$ & $-7.53$ & $3.27$ \\
        $28$ &	$38$&	$^{66}$Ni  & $8.23\cdot10^{-7}$  & $8.64\cdot10^{-7}$ &	$3.78\cdot10^{-7}$ & $8.048\cdot10^{-4}$ & $-7.5$  & $3.33$ \\
        $36$ &	$50$&	$^{86}$Kr  & $8.83\cdot10^{-7}$  & $1.87\cdot10^{-6}$ &	$1.03\cdot10^{-6}$ & $1.740\cdot10^{-3}$ & $-7.00$ & $4.26$ \\
        $34$ &	$50$&	$^{84}$Se  & $1.93\cdot10^{-6}$  & $6.81\cdot10^{-6}$ &	$5.59\cdot10^{-6}$ & $6.351\cdot10^{-3}$ & $-5.87$ & $6.46$ \\ 
        $32$ &	$50$&	$^{82}$Ge  & $7.06\cdot10^{-6}$  & $1.67\cdot10^{-5}$ &	$1.77\cdot10^{-6}$ & $1.560\cdot10^{-2}$ & $-4.81$ & $8.59$ \\
        $30$ &	$50$&	$^{80}$Zn  & $1.74\cdot10^{-5}$  & $3.81\cdot10^{-5}$ &	$5.05\cdot10^{-5}$ & $3.563\cdot10^{-2}$ & $-3.58$ & $11.15$\\
        $28$ &	$50$&	$^{78}$Ni  & $3.98\cdot10^{-5}$  &  &	 & \\ 
       \hline
        QMC$\pi$-III &             &           &                               &                              &                                 &                               \\
       \hline 
        $28$ &	$50$&	$^{78}$Ni  &                     & $5.05\cdot10^{-5}$ &	$6.96\cdot10^{-5}$ & $4.726\cdot10^{-2}$ & $-3.16$ & $12.08$ \\
        $44$ &	$82$&	$^{126}$Ru & $5.27\cdot10^{-5}$  & $7.57\cdot10^{-5}$ &	$1.13\cdot10^{-4}$ & $7.087\cdot10^{-2}$ & $-2.47$ & $13.67$ \\
        $42$ &	$82$&	$^{124}$Mo & $7.81\cdot10^{-5}$  & $1.00\cdot10^{-4}$ &	$1.59\cdot10^{-4}$ & $9.397\cdot10^{-2}$ & $-1.96$ & $14.87$\\
        $40$ &  $82$&	$^{122}$Zr & $1.04\cdot10^{-4}$  & $1.31\cdot10^{-4}$ &	$2.17\cdot10^{-4}$ & $0.122$  & $-1.46$ & $16.08$\\
        $38$ &	$82$&	$^{120}$Sr & $1.35\cdot10^{-4}$  & $2.03\cdot10^{-4}$ & $3.73\cdot10^{-4}$ & $0.190$  & $-0.53$ & $18.39$\\
        $36$ &	$82$&	$^{118}$Kr & $2.11\cdot10^{-4}$  & $2.61\cdot10^{-4}$ &	$4.95\cdot10^{-4}$ & $0.244$ & $\sim 0$
        & $19.74$\\
        \hline

        FRDM &             &                       &                               &                              &                                 &                               \\
   \hline
        $28$ &	$50$&	$^{78}$Ni  &                    & $5.22\cdot10^{-5}$ & $7.26\cdot10^{-5}$ & $4.877\cdot10^{-2}$ & $-3.1$  & $12.2$  \\ 
        $44$ &	$82$&	$^{126}$Ru & $5.44\cdot10^{-5}$ & $7.82\cdot10^{-5}$ & $1.18\cdot10^{-4}$ & $7.314\cdot10^{-2}$ & $-2.41$ & $13.82$\\
        $42$ &	$82$&	$^{124}$Mo & $8.05\cdot10^{-5}$ & $1.13\cdot10^{-4}$ & $1.85\cdot10^{-4}$ & $0.105$ & $-1.71$ & $15.45$ \\
        $40$ &	$82$&	$^{122}$Zr & $1.16\cdot10^{-4}$ & $1.60\cdot10^{-4}$ & $2.84\cdot10^{-4}$ & $0.150$ & $-0.99$ & $17.19$\\
        $38$ &	$82$&	$^{120}$Sr & $1.66\cdot10^{-4}$ & $2.26\cdot10^{-4}$ & $4.30\cdot10^{-4}$ & $0.212$ & $-0.23$ & $19.06$\\
        $36$ &	$82$&	$^{118}$Kr & $2.35\cdot10^{-4}$ & $2.57\cdot10^{-4}$ & $4.86\cdot10^{-4}$ & $0.241$ & $\sim 0$
        & $19.65$\\  
        \hline
        
         HFB24 &             &                       &                      &                              &                                 &                               \\
    \hline
        $28$ &	$50$&	$^{78}$Ni  &                    & $6.20\cdot10^{-5}$ &$9.14\cdot10^{-5}$ & $5.798\cdot10^{-2}$ & $-2.77$ & $17.23$\\ 
        $44$ &	$82$&	$^{126}$Ru & $6.27\cdot10^{-5}$ & $7.57\cdot10^{-5}$ & $1.13\cdot10^{-4}$ & $7.08\cdot10^{-2}$ & $-2.46$ & $18.23$\\
        $42$ &	$82$&	$^{124}$Mo & $7.81\cdot10^{-5}$ & $1.22\cdot10^{-4}$ & $2.06\cdot10^{-4}$ & $0.115$ & $-1.52$ & $21.17$\\
        $40$ &	$82$&	$^{122}$Zr & $1.26\cdot10^{-4}$ & $1.58\cdot10^{-4}$ & $2.77\cdot10^{-4}$ & $0.147$ & $-1.01$ & $22.79$\\
          $39$ &	$82$&	$^{121}$Y & $1.61\cdot10^{-4}$ & $1.65\cdot10^{-4}$ & $2.90\cdot10^{-4}$ & $0.154$ & $-0.94$ & $22.99$\\
        $38$ &	$82$&	$^{120}$Sr & $1.68\cdot10^{-4}$ & $1.95\cdot10^{-4}$ & $3.52\cdot10^{-4}$ & $0.182$ & $-0.59$ & $24.17$\\
        $38$ &	$84$&	$^{122}$Sr & $1.98\cdot10^{-4}$ & $2.38\cdot10^{-4}$ & $4.51\cdot10^{-4}$ & $0.224$ & $-0.13$ & $25.72$\\
        $38$ &	$86$&	$^{124}$Sr & $2.43\cdot10^{-4}$ & $2.55\cdot10^{-4}$ & $4.85\cdot10^{-4}$ & $0.239$ & $\sim 0$
        & $26.17$\\
        \hline
        
         NL3 &             &                       &                      &                              &                                 &                  & \\ 
 \hline
        $28$ &	$50$&	$^{78}$Ni  &                    & $6.28\cdot10^{-5}$ & $9.31\cdot10^{-5}$ & $5.877\cdot10^{-2}$ & $-2.74$ & $12.98$\\ 
        $44$ &	$82$&	$^{126}$Ru & $6.56\cdot10^{-5}$ & $7.19\cdot10^{-5}$ & $1.06\cdot10^{-4}$ & $6.73\cdot10^{-2}$  & $-2.56$ & $13.44$\\
        $42$ &	$82$&	$^{124}$Mo & $7.42\cdot10^{-5}$ & $9.21\cdot10^{-5}$ & $1.42\cdot10^{-4}$ & $8.82\cdot10^{-2}$  & $-2.12$ & $14.45$\\
        $40$ &	$82$&	$^{122}$Zr & $9.53\cdot10^{-5}$ & $1.26\cdot10^{-4}$ & $2.05\cdot10^{-4}$ & $0.118$ & $-1.54$ & $15.86$\\
        $38$ &	$82$&	$^{120}$Sr & $1.3\cdot10^{-4}$  & $1.65\cdot10^{-4}$ & $2.83\cdot10^{-4}$ & $0.155$ & $-1.01$ & $17.17$\\
        $36$ &	$82$&	$^{118}$Kr & $1.72\cdot10^{-4}$ & $2.32\cdot10^{-4}$ & $4.25\cdot10^{-4}$ & $0.218$ & $-0.3$ & $18.0$\\  
        $36$ &	$84$&	$^{120}$Kr & $2.37\cdot10^{-4}$ & $2.67\cdot10^{-4}$ & $5.00\cdot10^{-4}$ & $0.250$ & 
        $\sim 0$
        & $19.78$\\
        \hline
        \hline
	\end{tabular} 
	\egroup
\end{table*} 

\clearpage
 \begin{table*}[t!] 
 	\centering
 	\setlength{\tabcolsep}{7pt}
 	\def\arraystretch{0.95}
 	\caption{Results for neutron star models of 1.0, 1.4 and 1.94 $M_\odot$ calculated with the QMC$\pi$-III equation of state. The columns give the total (gravitational) mass $M_g$ and radius $R$, the core mass $M_c$ and radius $R_c$, and the mass and thickness of the inner and outer crust ($M_{in}$,$z_{in}$) and ($M_{out}$,$z_{out}$) respectively. All masses are in units of $M_\odot$ and radii are in km. \label{tab:comp}}
	\vspace{3mm}
 	\begin{tabular}{c c c c c c c c c}
 		\hline
 		\hline
    Model        &  M$_g$     & R        & M$_c$   & R$_c$    & M$_{in}$            & z$_{in}$ & M$_{out }$         & z$_{out}$ \\
 \hline   
   QMC$\pi$-III  & $1.0$  & $12.972$ & $0.951$ & $11.091$ & $0.486\cdot10^{-1}$ & $1.084$ & $0.650\cdot10^{-4}$ & $0.796$ \\
                 & $1.4$  & $13.008$ & $1.363$ & $11.753$ & $0.368\cdot10^{-1}$ & $0.739$ & $0.447\cdot10^{-4}$ & $0.517$ \\
                 & $1.94$ & $12.514$ & $1.919$ & $11.820$ & $0.208\cdot10^{-1}$ & $0.415$ & $0.233\cdot10^{-4}$ & $0.279$ \\ 
                         \hline
 	\end{tabular} 
\end{table*} 

\clearpage
 \begin{table*}[t!] 
 	\centering
 	\setlength{\tabcolsep}{5pt}
 	\def\arraystretch{0.65}
 	\caption{Properties of individual layers in the outer crust: Depth $z$ in km and mass $\Delta M$ (as a percentage of the total mass of the outer crust) are calculated for 1.0, 1.4 and 1.94  $M_\odot$ neutron stars using the QMC$\pi$-III, FRDM, HFB24 and NL3 mass models. \label{tab:layers}}
	\vspace{3mm}
 	\begin{tabular}{c c c c c c c}
 		\hline
 		\hline
 		 Element   & z$_{1.0}$           &  z$_{1.4}$         &  z$_{1.94}$      &  $\Delta M_{1.0} $  &  $\Delta M_{1.4} $ &  $\Delta M_{1.94} $ \\
 		\hline
 	    $^{56}$Fe  & $1.77\cdot10^{-2}$ & $1.13\cdot10^{-2}$ &$5.98\cdot10^{-3}$ &$8.51\cdot10^{-5}$  & $7.9 \cdot10^{-5}$ & $7.43\cdot10^{-5}$ \\
 		$^{62}$Ni  & $6.67\cdot10^{-2}$ & $4.25\cdot10^{-2}$ &$2.26\cdot10^{-2}$ &$1.07\cdot10^{-2}$  & $1.0\cdot10^{-2}$  & $9.51\cdot10^{-3}$ \\
 		$^{64}$Ni  & $6.66\cdot10^{-2}$ & $4.25\cdot10^{-2}$ &$2.27\cdot10^{-2}$ &$7.62\cdot10^{-2}$  & $7.16\cdot10^{-2}$ & $6.83\cdot10^{-2}$ \\
 		$^{66}$Ni  & $2.63\cdot10^{-3}$ & $1.68\cdot10^{-3}$ &$8.98\cdot10^{-4}$ &$5.87\cdot10^{-3}$  & $5.54\cdot10^{-3}$ & $5.3 \cdot10^{-3}$ \\
 		$^{86}$Kr  & $4.49\cdot10^{-2}$ & $2.88\cdot10^{-2}$ &$1.54\cdot10^{-2}$ &$1.58\cdot10^{-1}$  & $1.5\cdot10^{-1}$  & $1.43\cdot10^{-1}$ \\
 		$^{84}$Se  & $1.02\cdot10^{-1}$ & $6.54\cdot10^{-2}$ &$3.51\cdot10^{-2}$ &$1.07$              & $1.02$             & $0.98$             \\
 		$^{82}$Ge  & $9.36\cdot10^{-2}$ & $6.06\cdot10^{-2}$ &$3.27\cdot10^{-2}$ &$2.75$              & $2.65$             & $2.59$             \\
 		$^{80}$Zn  & $1.06\cdot10^{-1}$ & $6.89\cdot10^{-2}$ &$3.73\cdot10^{-2}$ &$7.2$               & $7.03$             & $6.9$             \\
 \hline
        QMC$\pi$-III  &                 &                    &                    &         &         &         \\
 \hline
        $^{78}$Ni  & $3.60\cdot10^{-2}$ & $2.35\cdot10^{-2}$ & $1.28\cdot10^{-2}$ & $4.07$  & $4.0$   & $3.95$  \\
  		$^{126}$Ru & $5.75\cdot10^{-2}$ & $3.77\cdot10^{-2}$ & $2.05\cdot10^{-2}$ & $9.14$  & $9.05$  & $8.97$  \\
 		$^{124}$Mo & $4.27\cdot10^{-2}$ & $2.8 \cdot10^{-2}$ & $1.53\cdot10^{-2}$ & $9.39$  & $9.34$  & $9.31$  \\
 		$^{122}$Zr & $4.13\cdot10^{-2}$ & $2.72\cdot10^{-2}$ & $1.49\cdot10^{-2}$ & $11.9$  & $11.9$  & $11.9$  \\
 		$^{120}$Sr & $7.57\cdot10^{-2}$ & $5.0 \cdot10^{-2}$ & $2.74\cdot10^{-2}$ & $30.96$ & $31.18$ & $31.35$ \\
 		$^{118}$Kr & $4.01\cdot10^{-2}$ & $2.66\cdot10^{-2}$ & $1.46\cdot10^{-2}$ & $22.64$ & $22.96$ & $23.2 $ \\
 \hline
        FRDM       &                    &                    &                    &         &         &         \\
 \hline
        $^{78}$Ni  & $4.09\cdot10^{-2}$ & $2.68\cdot10^{-2}$ & $1.45\cdot10^{-2}$ & $4.73$  & $4.65$  & $4.6$   \\
 		$^{126}$Ru & $5.81\cdot10^{-2}$ & $3.81\cdot10^{-2}$ & $2.07\cdot10^{-2}$ & $9.57$  & $9.47$  & $9.4$   \\
 		$^{124}$Mo & $5.8 \cdot10^{-2}$ & $3.82\cdot10^{-2}$ & $2.08\cdot10^{-2}$ & $13.83$ & $13.78$ & $13.75$ \\
 		$^{122}$Zr & $5.9 \cdot10^{-2}$ & $3.9 \cdot10^{-2}$ & $2.13\cdot10^{-2}$ & $20.0$  & $20.08$ & $20.13$ \\
 		$^{120}$Sr & $6.1 \cdot10^{-2}$ & $4.04\cdot10^{-2}$ & $2.22\cdot10^{-2}$ & $29.02$ & $29.32$ & $29.57$ \\
 		$^{118}$Kr & $1.81\cdot10^{-2}$ & $1.2 \cdot10^{-2}$ & $6.61\cdot10^{-3}$ & $10.79$ & $10.96$ & $11.09$ \\
 \hline
 	    HFB24      &                    &                    &                    &         &         &         \\
 \hline
        $^{78}$Ni  & $6.9 \cdot10^{-2}$ & $4.51\cdot10^{-2}$ & $2.45\cdot10^{-2}$ & $8.8$   & $8.68$  & $8.58$  \\
 		$^{126}$Ru & $2.58\cdot10^{-2}$ & $1.69\cdot10^{-2}$ & $9.22\cdot10^{-2}$ & $4.55$  & $4.51$  & $4.48$  \\
 		$^{124}$Mo & $7.85\cdot10^{-2}$ & $5.17\cdot10^{-2}$ & $2.82\cdot10^{-2}$ & $19.41$ & $19.36$ & $19.32$ \\
 		$^{122}$Zr & $4.12\cdot10^{-2}$ & $2.72\cdot10^{-2}$ & $1.49\cdot10^{-2}$ & $14.44$ & $14.5$  & $14.55$ \\
 		$^{121}$Y  & $4.72\cdot10^{-2}$ & $3.12\cdot10^{-2}$ & $1.71\cdot10^{-2}$ & $1.89$  & $1.9$   & $1.91$  \\
 		$^{120}$Sr & $2.86\cdot10^{-2}$ & $1.89\cdot10^{-2}$ & $1.04\cdot10^{-2}$ & $12.72$ & $12.83$ & $12.92$ \\
 		$^{122}$Sr & $3.67\cdot10^{-2}$ & $2.43\cdot10^{-2}$ & $1.33\cdot10^{-2}$ & $19.57$ & $19.82$ & $10.01$ \\
 		$^{124}$Sr & $1.0 \cdot10^{-2}$ & $6.66\cdot10^{-3}$ & $3.66\cdot10^{-3}$ & $6.09$  & $6.18$  & $6.26$  \\
 \hline
        NL3       &                    &                    &                    &         &         &         \\
 \hline
        $^{78}$Ni  & $7.12\cdot10^{-2}$ & $4.66\cdot10^{-2}$ & $2.53\cdot10^{-2}$ & $8.99$  & $8.87$  & $8.77$   \\
 		$^{126}$Ru & $1.5\cdot10^{-2}$  & $9.84\cdot10^{-2}$ & $5.36\cdot10^{-2}$ & $2.55$  & $2.53$  & $2.51$   \\
 		$^{124}$Mo & $3.61\cdot10^{-2}$ & $2.38\cdot10^{-2}$ & $1.3 \cdot10^{-2}$ & $7.39$  & $7.35$  & $7.31$   \\
 		$^{122}$Zr & $4.83\cdot10^{-2}$ & $3.18\cdot10^{-2}$ & $1.74\cdot10^{-2}$ & $13.03$ & $13.01$ & $13.0$   \\
 		$^{120}$Sr & $4.28\cdot10^{-2}$ & $2.82\cdot10^{-2}$ & $1.55\cdot10^{-2}$ & $15.32$ & $15.39$ & $15.43$  \\
 		$^{118}$Kr & $5.75\cdot10^{-2}$ & $3.81\cdot10^{-2}$ & $2.09\cdot10^{-2}$ & $27.87$ & $28.15$ & $28.37$  \\
 		$^{120}$Kr & $2.14\cdot10^{-2}$ & $1.42\cdot10^{-2}$ & $7.8 \cdot10^{-3}$ & $12.83$ & $13.01$ & $13.16 $ \\
 \hline
 \hline
 	\end{tabular} 
\end{table*} 

\clearpage
 \begin{table*}[t!] 
 	\centering
 	\setlength{\tabcolsep}{10pt}
 	\def\arraystretch{0.65}
 	\caption{Properties of the individual outer crust layers for a $1.4\, M_{\odot}$ neutron star calculated with the QMC$\pi$-III mass model using measured masses where available but otherwise masses predicted using QMC$\pi$-III. The ion density, $n_b$, electron density, $n_e$, electron specific heat capacity, $C_e$, and electron Fermi momentum, $p_F^e$, are given for each layer.}
	\label{tab:C_e}
	\vspace{3mm}
 	\begin{tabular}{c c c c c }
 		\hline
 		\hline
 		 Element   & $n_b$          &  $n_e$         &  $C_e$                 & p$_F^e$\\
 		           & [fm$^{-3}$]  & [fm$^{-3}$]    & [MeV fm$^{-3}$ K$^{-1}$] & [ MeV ]   \\
 		\hline
 	    $^{56}$Fe   & $0.32\cdot10^{-10}$  & $0.68\cdot10^{-10}$ & $0.20\cdot10^{-22}$  & $0.42\cdot10^{-7}$ \\
 		$^{62}$Ni   & $0.97\cdot10^{-9}$   & $0.21\cdot10^{-8}$  & $0.19\cdot10^{-21}$  & $0.13\cdot10^{-5}$ \\
 		$^{64}$Ni   & $0.67\cdot10^{-8}$   & $0.15\cdot10^{-7}$  & $0.74\cdot10^{-21}$  & $0.94\cdot10^{-5}$\\
 		$^{66}$Ni   & $0.13\cdot10^{-7}$   & $0.30\cdot10^{-7}$  & $0.12\cdot10^{-20}$  & $0.18\cdot10^{-4}$\\
 		$^{86}$Kr   & $0.16\cdot10^{-7}$   & $0.37\cdot10^{-7}$  & $0.13\cdot10^{-20}$  & $0.23\cdot10^{-4}$\\
 		$^{84}$Se   & $0.48\cdot10^{-7}$   & $0.12\cdot10^{-6}$  & $0.29\cdot10^{-20}$  & $0.72\cdot10^{-4}$\\
 		$^{82}$Ge   & $0.14\cdot10^{-6}$   & $0.36\cdot10^{-6}$  & $0.60\cdot10^{-20}$  & $0.22\cdot10^{-3}$\\
 		$^{80}$Zn   & $0.34\cdot10^{-6}$   & $0.89\cdot10^{-6}$  & $0.11\cdot10^{-19}$  & $0.55\cdot10^{-3}$\\
 	    $^{78}$Ni     & $0.58\cdot10^{-6}$   & $0.16\cdot10^{-5}$  & $0.16\cdot10^{-19}$  & $0.98\cdot10^{-3}$\\
 \hline
        QMC$\pi$-III  &            &                &           & \\
 \hline
  		$^{126}$Ru    & $0.51\cdot10^{-6}$   & $0.15\cdot10^{-5}$  & $0.15\cdot10^{-19}$  & $0.89\cdot10^{-3}$\\
 		$^{124}$Mo    & $0.72\cdot10^{-6}$   & $0.21\cdot10^{-5}$  & $0.20\cdot10^{-19}$  & $0.13\cdot10^{-2}$\\
 		$^{122}$Zr    & $0.96\cdot10^{-6}$   & $0.29\cdot10^{-5}$  & $0.25\cdot10^{-19}$  & $0.18\cdot10^{-2}$\\
 		$^{120}$Sr    & $0.14\cdot10^{-5}$   & $0.44\cdot10^{-5}$  & $0.32\cdot10^{-19}$  & $0.27\cdot10^{-2}$\\
 		$^{118}$Kr    & $0.20\cdot10^{-5}$   & $0.65\cdot10^{-5}$  & $0.42\cdot10^{-19}$  & $0.40\cdot10^{-2}$\\
 \hline
 \hline
 	\end{tabular} 
\end{table*} 

\clearpage

\clearpage

\clearpage

\clearpage

\end{document}